\documentclass[hyper]{JHEP} 

\usepackage{epsfig}
\usepackage{amsfonts}
\usepackage{graphicx}

\def\be{\begin{equation}}
\def\ee{\end{equation}}
\def\ba{\begin{array}}
\def\ea{\end{array}}
\def\bea{\begin{eqnarray}}
\def\eea{\end{eqnarray}}
\def\nn{\nonumber\\}

\def\ba{\mathbf{a}}

\def\a{\alpha}

\def\ph{\phi}

\def\l{\lambda}

\def\th{\theta}

\def\r{\rho}

\def\pr{\prime}
\def\tR{\tilde{R}}

\def\lb{\left[}

\def\ls{\left(}

\def\rb{\right]}

\def\rs{\right)}

\def\det{{\rm det}}

\newcommand\fverb{\setbox\pippobox=\hbox\bgroup\verb}

\newcommand\fverbdo{\egroup\medskip\noindent%

            \fbox{\unhbox\pippobox}\ }

\newcommand\fverbit{\egroup\item[\fbox{\unhbox\pippobox}]}

\newbox\pippobox

\title{ Wilson Loops from D-branes in AdS$_4
\times {\bf CP}^3$ with $B_{NS}$ Holonomy}
\author{J. Kluso\v{n}\\
Department of
Theoretical Physics and Astrophysics\\
Faculty of Science, Masaryk University\\
Kotl\'{a}\v{r}sk\'{a} 2, 611 37, Brno\\
Czech Republic\\

E-mail: \email{klu@physics.muni.cz}}

\author{Kamal L. Panigrahi\\
Department of Physics \& Meteorology and \\
Center for Theoretical Studies \\
Indian Institute of Technology Kharagpur, \\
Kharagpur, 721 302, INDIA \\
E-mail: \email{panigrahi@phy.iitkgp.ernet.in}}

\preprint{{0906.2148[hep-th]}}

\abstract{ We study Wilson loops in N=6 superconformal Chern-Simons theory with gauge group
$U(M) \times \overline{U(N)}$ that is dual to N M2-branes and (M-N) fractional
M2-branes, equivalently, discrete 3-form holonomy at C4/Zk orbifold singularity.
We give description of these Wilson loops in terms of macroscopic fundamental string
and D6-branes in the dual AdS$_4\times$ CP$^3$ geometry with B$_{\rm{NS}}$ holonomy turned on
over CP$^1 \subset$ CP$^3$.}

\def \ba{\mathbf{a}}
\def\tr{\mathrm{Tr}}

\def\bA{\mathbf{A}}

\def\mF{\mathcal{F}}

\def\mL{\mathcal{L}}

\def\tR{\tilde{R}}

\begin{document}

\section{Introduction and Summary}\label{first}
The ABJM \cite{Aharony:2008ug} theory has been conjectured to be
dual to M-theory on AdS$_4 \times S^7/Z_k$ with $N$ units of
four-form flux which for $k << N << k^5$ can be compactified to
type IIA theory on AdS$_4 \times CP^3$, where $k$ is the level of
Chern-Simon theory with gauge group $U(N) \times U(N)$. The ABJM theory is
weakly coupled for $\lambda << 1$, where $\lambda = N/k$ is the 't
Hooft coupling. In continuation with this proposal, Aharony,
Bergman, and Jafferis (ABJ) \cite{Aharony:2008gk} identified a
further class of gauge-gravity duality with extended supersymmetry
namely, a three dimensional ${\cal N}$ = 6 superconformal
Chern-Simons theory with a gauge group $U(M)_k \times
{\overline{U(N)}}_{-k}$, with $k$ being the level of the CS
theory, is dual to type IIA string theory on AdS$_4 \times {\bf
CP}^3$ with B$_{NS}$ holonomy turned on over {\bf CP}$^1$
$\subset$ {\bf CP}$^3$ \footnote{For further study of ABJ theory,
see
\cite{Bak:2008vd,Kluson:2009tz,Caputa:2009ug,Bergman:2009zh,Bagger:2008se,Jain:2008mt,Fujita:2009nm}.}.
In gauge theory, Wilson loop operators are non-local gauge
invariant operators in which the theory can be formulated. One
defines a Wilson loop as the trace in an arbitrary representation
$R$ of the gauge group $G$ of the holonomy matrix associated with
parallel transport along a closed curve $C$ in spacetime. Since
the beginning of the proposed AdS/CFT correspondence
\cite{Maldacena:1997re}, it is known that Wilson loops in $N=4$
SYM theory can be calculated in dual description using macroscopic
strings \cite{Maldacena:1998im,Rey:1998ik}. In a recent
interesting paper \cite{Ishizeki:2008dw} a class of  new open
strings in $AdS_5$ were found arising out of the solutions to the
equations of motion corresponding to fundamental strings and they
describe Wilson loops in the fundamental representations. Some
time back it was argued in very interesting paper
\cite{Gomis:2006sb}, for type IIB theory, that Wilson loops have a
gravitational dual description in terms of D5-branes or
alternatively in terms of D3-branes in $AdS_5\times S^5$
background \footnote{For closely related works, see
\cite{Bonelli:2008rv,Chen:2008ds,Drukker:2007qr,Lunin:2007zz,Sakaguchi:2007ea,
Chu:2007pb,Chen:2007ir,Gomis:2006mv,Drukker:2006zk,Gomis:2006im,
Armoni:2006ux,Hartnoll:2006ib,Drukker:2006ga,Lunin:2006xr,Hartnoll:2006hr}.}.

Recently the Wilson loop operators have been investigated in the
three-dimensional, N=6 superconformal Chern-Simons theory dual to
IIA superstring theory on AdS$_4 \times$ CP$^3$ in
\cite{Drukker:2008zx,Chen:2008bp,Rey:2008bh,Kluson:2008wn}. For a
given contour there are two linear combinations of Wilson loop
operators transforming oppositely under time-reversal
transformation. Like the AdS$_5 \times$ S$^5$ case the D-brane
configurations wrapping various cycles in the CP$^3$ has been
shown to correspond to dual operators in the boundary theory. We
generalize these to the case of ABJ theory, which has an
additional B$_{NS}$ holonomy turned on over CP$^1 \subset$ CP$^3$.
We propose Wilson loops containing both gauge potential and a pair
of bi-fundamental fields almost in the same way as in case of ABJM
theory. However there is a subtle difference that now the gauge
groups are $U(M)$ and $U(N)$ respectively. Further, the functional
form of the Wilson loop operator is constrained by the requirement
of affine symmetry along the contour $C$ and by superconformal
symmetry $R^{1,2}$ and by gauge and $SU(4)$ symmetries. We
generalize the analysis given in \cite{Rey:2008bh} and find out
that there are two elementary Wilson loops operators. We further
give Wilson loop operators in terms of objects in the dual
supergravity on AdS$_4 \times$ CP$^3$ with B$_{\rm NS}$ holonomy.

Firstly, the fundamental string description of Wilson loop is
precisely the same as in case of ABJM theory and hence again these
strings describe symmetric combinations of Wilson loops in
fundamental representation. As it will be clear in what follows
that the effect of the non-zero torsion is sub-leading in the
supergravity description. Equivalently, the dynamics of the string
with world-volume topology $AdS_2 \subset AdS_4$ is not affected
by the NS-NS two form field. For the D2-brane description of
Wilson loop that span the subspace $AdS_2\times S^1 \subset AdS_4
\times CP^3$, the presence of two form $B_{NS}$ does not affect
this. This solution was studied in \cite{Kluson:2008wn}. We study
the Wilson loop solution corresponding to D6-brane with topology
$AdS_2\times\Sigma_4$ where $\Sigma_4 \subset$ CP$^3$. More
precisely we consider D6-brane configuration that spans $AdS_2$
subspace of $AdS_4$ and couples to $C_{5}$. We find out that this
D6-brane describes the Wilson loops in anti-symmetric
representation.

The rest of the paper is organized as follows. In section-2 we
review basic facts about the ABJ theory. Section-3 is devoted to
review of basic properties of Wilson loops in the supersymmetric
$U(M) \times \overline{U(N)}$ gauge theory. In section-4 we study
these Wilson loops from the supergravity dual of AdS$_4 \times$
CP$^3$ with B$_{\rm NS}$ holonomy and suggest their relations to the
Wilson loops in ABJ theory.
\section{Review of ABJ Theory}\label{second}
In this section we review basic facts about ABJ theory. This
theory is a $N=6$ supersymmetric Chern-Simons  matter theory with
two gauge groups of ranks $N,M$ and levels $k$ and $-k$
respectively. In addition to the gauge fields there are bosonic
and fermionic field $Y^A$ and $\psi_A$ respectively. More
precisely this theory has the following properties:
\begin{itemize}
\item Gauge and global symmetries: \vskip 4mm \quad gauge
symmetry: \ $U(M)\times \overline{U(N)} $ \vskip 4mm \quad global
symmetry: \ $SU(4)$ \item The field content of given theory is as
follows:
\begin{eqnarray}
& &A_\mu^{(L)}: \ \mathrm{Adj}(U(M)) \ , \quad  A_\mu^{(R)}: \
\mathrm{Adj}(\overline{U(N)})
 \ . \nonumber \\
\end{eqnarray}
Further there is $M\times N$ matrix valued matter fields-$4$
complex scalar $Y^A (A=1,2,3,4)$ and their hermitian conjugates
$Y^\dag_A$.
There are also $M\times N$ matrix valued  fermions $\psi_A$
together with their hermitian conjugates $\psi^{A\dag}$. Fields
with superscript $A$ index transform in the $\mathbf{4}$ of $R$
symmetry $SU(4)$ group and those with subscript index transform in
the $\overline{\mathbf{4}}$ representations.
\end{itemize}
%
The corresponding Lagrangian has the
following form
\footnote{Note that we use the notation as in
\cite{Bandres:2008ry}.}
\begin{eqnarray}\label{Act}
\mL &=&-\frac{k}{2\pi}\tr (D^\mu
Y^\dag_A D_\mu Y^A)+\frac{ik}{2\pi} \tr
(\overline{\psi}^{A\dag} \gamma^\mu
D_\mu\psi_A)-V+\mL_{CS}+
\nonumber \\
&+& i\tr (\overline{\psi}^{A\dag}\psi_A
Y_B^\dag Y^B)-i\tr (\psi_A
\overline{\psi}^{\dag A}Y^BY_B^\dag)+
\nonumber \\
&+& 2i \tr(\psi_A
\overline{\psi}^{B\dag}Y^A Y_B^\dag)
-2i\tr(\overline{\psi}^B\psi_AY^\dag_BY^A)+
\nonumber \\
&+& i \epsilon^{ABCD}\tr(\psi_A
Y^\dag_B \psi_C Y^\dag_D)-
  i
 \epsilon_{ABCD} \tr
 (\overline{\psi}^{A\dag}Y^B
\overline{\psi}^{C\dag}Y^D) \ ,
  \nonumber \\
 \end{eqnarray}
where $\mL_{CS}$ is a Chern-Simons term
and $V(Y)$ is a sextic scalar potential
\begin{eqnarray}\label{CSact}
\mL_{CS}&=&\frac{k}{4\pi}
\epsilon^{\mu\nu\lambda}
\tr[A^{(L)}_\mu
\partial_\nu A_\lambda^{(L)}+
\frac{2i}{3}A_\mu^{(L)}A_\nu^{(L)}A_\lambda^{(L)}]
-\nonumber \\
&-&\frac{k}{4\pi} \epsilon^{\mu\nu\lambda}
\overline{\tr}[A_\mu^{(R)}\partial_\nu
A_\lambda^{(R)}+\frac{2i}{3}A_\mu^{(R)}
A_\nu^{(R)}A_\lambda^{(R)}] \ ,
\nonumber \\
V(Y)&=& -\frac{1}{3}\tr [ Y^AY_A^\dag
Y^B Y_B^\dag Y^C Y_C^\dag +Y_A^\dag Y^A
Y_B^\dag Y^B Y_C^\dag
Y^C+\nonumber \\
&+& 4 Y^AY_B^\dag Y^C Y_A^\dag Y^B
Y_C^\dag -6 Y^A Y_B^\dag Y^B Y_A^\dag
Y^C Y_C^\dag] \ . \nonumber \\
\end{eqnarray}
Further,   the covariant derivatives for the scalars are defined
as
\begin{eqnarray}\label{Dcov}
D_\mu Y^A=\partial_\mu Y^A
+iA^{(L)}_\mu Y^A-iY^A
A^{(R)}_\mu \ , \quad
D_\mu Y^\dag_A=
\partial_\mu Y^\dag_A -iY^\dag_A
A_\mu^{(L)}+i A_\mu^{(R)}Y^\dag_A
 \ , \nonumber \\
\end{eqnarray}
and for fermions
\begin{eqnarray}
D_\mu \psi_A=
\partial_\mu \psi_A+i
A^{(L)}_\mu \psi_A-i\psi_A A_\mu^{(R)}
\ , \quad D_\mu \psi^{\dag A}=
\partial_\mu \psi^{A\dag}
-i\psi^{\dag A}A_\mu^{(L)}+ i
A_\mu^{(R)}\psi^\dag_A \ .  \nonumber \\
\end{eqnarray}
In sharp contrast from $N=4$ SYM theory, well known from the
$AdS_5/CFT_4$ correspondence, the characteristic property of ABJ
theory is that the coupling parameters are integer valued. This
fact suggests that these coupling parameters cannot possibly run
under renormalization group flow. Then it is convenient to
introduce the parameters $\lambda,\overline{\lambda}$ and consider
the generalized t'Hooft planar limit
\begin{equation}
M,N,k\rightarrow \infty \quad \mathrm{with} \ \lambda\equiv
\frac{N}{k} \ , \quad \overline{\lambda}\equiv \frac{M}{k} \ ,
\quad b=\frac{(M-N)}{k} \quad \equiv \mathrm{fixed} \ .
\end{equation}
As was shown in \cite{Aharony:2008gk}
the number of fractional branes $M-N$
is limited to $0\leq (M-N) \leq k$.
This is an important fact since in the
planar limit this implies that
$b\in[0,1]$.  Let us discuss the parity
transformation in dual theory. By
definition the  parity transformation
maps the Chern-Simons parameters by $k$
to $-k$ while holding $M,N$ fixed. In
case of ABJM theory the parity
transformations  are defined as
\begin{equation}
\mathcal{P}: \quad t\rightarrow t,
\quad x^{1}\rightarrow x^{1} \quad x^2
\rightarrow - x^2, \quad
(A_\mu^{(L)},A_\mu^{(R)},Y^A,Y_A^\dag)
\rightarrow (-A_\mu^{(R)},-A_\mu^{(L)},
Y_A^\dag,Y^A) \ .
\end{equation}
Since this parity exchanges $Y^A$ with $Y_A^\dag$ it exchanges two
isomorphic groups $U(N)$ and $\overline{U(N)}$. It is also clear
that the ABJM theory is manifestly invariant under this parity
transformation. On the other hand in the ABJ theory the parity
transformation cannot be a symmetry of the theory since the two
gauge groups are different and cannot be exchanged. On the other
hand as nicely argued in \cite{Aharony:2008gk} and in
\cite{Bak:2008vd} the parity transformation maps one ABJ theory
with a given gauge group to another ABJ theory with a different
gauge group. Explicitly, it was argued that there is an
equivalence relations between these two ABJ theories:
\begin{equation}
U(M)_k \times \overline{U(N)}_{-k}
\simeq U(N)_k \times
\overline{U(M)}_{-k} \ .
\end{equation}
On the other hand we can expect that at the  strong coupling limit
 $(\lambda,\overline{\lambda}>> 1)$, where the supergravity
description is effective, the effects that are parameterized by
$b$ could be invisible since $b << \lambda,\overline{\lambda}$.
However we will see in the next section that in case of D6-brane
description of the Wilson loops the vacuum expectation value of
the Wilson loop in anti-symmetric representation is sensitive to
the parameter $b$.

\section{Wilson loops in $N=6$
Superconformal Chern-Simons Theory} Our goal is to find Wilson
loop operators in ABJ theory. The proposed Wilson loops contain
both gauge potential and a pair of bi-fundamental fields almost in
the same way as in ABJM theory. However there is a subtle
difference that now the gauge groups are $U(M)$ and $U(N)$
respectively. Further, it was carefully discussed in
\cite{Rey:2008bh} that
 the
functional form of the Wilson loop
operator is constrained
 by the requirement of affine symmetry
along the contour $C$,  by superconformal symmetry along
$R^{1,2}$, and by gauge and $SU(4)$ symmetries. On the other hand
since these Wilson loops are independent it is clear that the
Wilson loops in ABJ theory have to obey the same conditions as
that of ABJM theory \cite{Rey:2008bh}.

Let us be more specific. We denote the coordinates of $R^{1,2}$ as
$x^\mu$ and of $SU(4)$ internal space as $z^I,\overline{z}_I$.
With two gauge fields $A^{(L)}_\mu$ of $U(M)$ and $A^{(R)}_\mu$ of
$\overline{U(N)}$ we can construct two types of Wilson loops
operators that are associated with each gauge groups. Let us
consider $U(M)$ gauge group. Following \cite{Rey:2008bh} we
propose the $U(M)$ Wilson loop operator as
\begin{equation}
W_M[C,P]=\frac{1}{M} \tr \mathcal{P}
\exp i\int_C d\tau (\dot{x}^\mu
A_\mu(x)+ M_A^B Y^A(x) Y^\dag_B(x)) \ ,
\end{equation}
where
 the vector field $\dot{x}^\mu(\tau)$
specifies the path $C$ in $R^{1,2}$ and
$M_A^{\ B}$ was defined in
\cite{Rey:2008bh}. Finally, the $\tr$
means the trace over $U(M)$ group.

In the same way we can propose the
Wilson loop operator of
$\overline{U(N)}$ in the form
\begin{equation}
\overline{W}_N[C,P]= \frac{1}{N}
\overline{\tr} \mathcal{P} \exp i
\int_C d\tau (\dot{x}^\mu
A_\mu^{(R)}(x)+ M_A^{\
B}(\tau)Y_B^\dag(x)Y^A(x))  \ ,
\end{equation}
where now $\overline{\tr}$ denotes the
trace over $\overline{U(N)}$ group.
From the point of view of ABJ theory more Wilson loops operators
are possible, for example we can consider linear combination
\begin{equation}
W_M[C,P]+\overline{W}_N[C,P]
\end{equation}
or the product of Wilson loop operators
\begin{equation}
W_N[C,P]\cdot \overline{W}_N[C,P]
\end{equation}
As was also nicely shown in
\cite{Rey:2008bh} the form of Wilson
loops is restricted by requirement that
they preserve some fractions of $N=6$
supersymmetry. However it turns out
that this does not determine possible
combinations of Wilson loops since
supersymmetry acts on $W_M[C,P]$ and
$\overline{W}_N[C,P]$ independently.

The fact that the  ABJ theory possesses two sets of Wilson loops,
one for $U(M)$ and the second one for $\overline{U(N)}$ the dual
holographic theory is puzzling. Explicitly, as in case of
$AdS_5/CFT_4$ correspondence we expect that the Wilson loops in
fundamental representation corresponds the fundamental string in
the bulk \cite{Maldacena:1998im,Rey:1998ik}. In case of ABJM
theory this problem was resolved in
\cite{Berenstein:2008dc,Drukker:2008zx,Rey:2008bh} where it was
argued that fundamental string corresponds to the symmetric
combinations of Wilson loops in dual theory. The question is
whether this picture changes in case of ABJ theory when the two
gauge groups are different and when the dual background possesses
discrete $B_{NS}$ holonomy. In particular, we would like to see
whether dual description is sensitive to the parity breaking
effects that are represented by the presence of non-trivial
$B_{NS}$ holonomy.
\section{Description of Wilson loops in
Supergravity Dual $AdS_4\times CP^3$ with $B_{NS}$ Holonomy} In
this section we try to find description of the Wilson loop
introduced in previous section in terms of macroscopic objects in
dual AdS$_4\times$ CP$^3$ geometry.

We start by writing down the metric for AdS$_4 \times$ CP$^3$,
which in a particular parametrization reads \bea
ds^2 &=& \tR^2( ds_{AdS_4}^2+4 ds_{CP_3}^2) \ , \nonumber \\
  ds_{AdS_4}^2 &=& \lb - \cosh^2 \r \
dt^2 + d\r^2 + \sinh^2 \r \ls d \th^2 +
\sin^2 \th d \ph^2 \rs \rb  \nn
ds_{CP_3}^2&=& \frac{1}{4} \left[
d\alpha^2 + \cos^2 \frac{\alpha}{2}
(d\theta_1^2+\sin^2\theta_1 d\varphi_1^2)+
\sin^2\frac{\alpha}{2}
(d\theta_2^2+\sin^2\theta_2 d\varphi_2^2)+
\right. \nonumber \\
&+&\left.\sin^2\frac{\alpha}{2}\cos^2\frac{\alpha}{2}
(d\chi+\cos\theta_1
d\varphi_1-\cos\theta_2 d\varphi_2)^2 \right] \ ,
\nonumber \\
\end{eqnarray}
where
\begin{equation}
0\leq \alpha,\theta_1,\theta_2 \leq \pi
\ , \quad 0\leq \varphi_1,\varphi_2\leq 2\pi
\ , \quad 0\leq \chi  \leq 4 \pi \ ,
\end{equation}
and where
\begin{equation}
\tR^2=\frac{R^3}{4k} \ , \quad
e^{2\Phi_0}=\frac{R^3}{k^3} \ .
\end{equation}
While taking the limit $\a^{\pr} =1$, the curvature radius $R$ is
given by $\tR^2 = \pi\sqrt{2\l} $. The 't Hooft coupling
constant is $\l \equiv N/k$ where $k$ is the level of the
3-dimensional N=6 ABJM model.

Let us now consider Ramond-Ramond
gauge fields. The first non-trivial
one is the one-form potential
$C_1$ that takes the form
\begin{equation}
C_1=\frac{k}{4} [\cos\alpha d\chi+2
\cos^2\frac{\alpha}{2}\cos\theta_1
d\varphi_1+2\sin^2\frac{\alpha}{2}\cos\theta_2
d\varphi_2] \ .
\end{equation}
Further there is a three-form
potential $C_3$ in the form
\begin{eqnarray}
C_3&=&\frac{1}{8}R^3\cosh^3 u \cosh \rho
dt\wedge d\rho\wedge d\phi \ , \quad
 \mathrm{appropriate \  for \  a  \ time-like \  curve} \ ,
 \nonumber \\
 C_3&=&\frac{1}{8}R^3 \cosh^3u \sinh\rho
 d\psi\wedge d\rho\wedge d\phi \ ,
 \quad
 \mathrm{appropriate \ for \ a  \  space -like  \ curve} \ .
 \nonumber \\
 \end{eqnarray}
 The Hodge dual of $F_4=dC_3$ is a six form
 $F_6$ that is
 proportional to the  volume form of $CP^3$
 \begin{equation}
F_6=\star F_4=
\frac{3R^6}{2^8 k}\sin^3\alpha
\sin\theta_1 \sin\theta_2  d\alpha
\wedge d\theta_1\wedge d\theta_2
\wedge d\chi \wedge d\varphi_1 \wedge \varphi_2 \ .
\end{equation}
Then we can define the five form $C_5$ as
\begin{equation}
C_5=-\frac{R^6}{2^8 k}(
\sin^2\alpha\cos\alpha+2\cos\alpha-2)
\sin\theta_1\sin\theta_2
d\theta_1\wedge d\theta_2 \wedge d\chi
\wedge d\varphi_1 \wedge d\varphi_2 \
\end{equation}
that obeys the relation $F_6=dC_5$.
The characteristic property of
 the background dual
to ABJ theory is an existence of an additional non-trivial NS-NS
two form
\begin{eqnarray}
B_{NS}&=&
-\frac{b}{4}
\sin \alpha d\alpha \wedge
(d\chi+\cos\theta_1 d\varphi_1-
\cos\theta_2 d\varphi_2)-\nonumber \\
&-& \frac{b}{2}\left(\cos^2 \frac{\alpha}{2}
\sin \theta_1 d\theta_1 \wedge
d\varphi_1 + \sin^2 \frac{\alpha}{2}
\sin \theta_2 d\theta_2 \wedge
d\varphi_2\right) \ .  \nonumber \\
 \end{eqnarray}
 Let us now consider  Dp-brane  in
 given background. Generally  the low energy
 dynamics   of Dp-brane  is governed by
following action
\begin{eqnarray}\label{actD1}
S&=&S_{DBI}+S_{WZ} \ , \nonumber \\
S_{DBI}&=&-\tau_p \int d^{p+1}\xi e^{-\Phi}
\sqrt{-\det\bA} \ , \nonumber \\
\bA_{\alpha\beta}&=&\partial_\alpha x^M\partial_\beta x^N
G_{MN}+2\pi\mF_{\alpha\beta}  \ , \nonumber \\
  \mF_{\alpha\beta}&=&
\partial_\alpha A_\beta-\partial_\beta A_\alpha-
(2\pi)^{-1}B_{MN}\partial_\alpha x^M\partial_\beta x^N
 \ , \nonumber \\
S_{WZ}&=&\tau_p\int e^{2\pi\mF}\wedge C \ ,  \nonumber \\
\end{eqnarray}
where $\tau_p$ is D$p$-brane tension,
$\xi^\alpha,\alpha=0, 1, \dots,p$ are
the $(p+1)$ world-volume coordinates
and where $A_\alpha$ is gauge field
living on the world-volume of
D$p$-brane. Further $\Phi, G_{MN}$ and
$B_{MN}$ are space-time dilaton,
graviton and NS two form field
respectively. Finally $C$ in the last
line in (\ref{actD1}) means collection
of Ramond-Ramond fields. In order to
study Wilson lines we perform an
analytic continuation when we introduce
$\tau$ as $t=i\tau$ so that the
Euclidean form of Dp-brane
 action takes the form
\begin{eqnarray}\label{actD1e}
S&=&S_{DBI}+S_{WZ} \ , \nonumber \\
S_{DBI}&=&\tau_p \int d^{p}\xi d\tau e^{-\Phi_0}
\sqrt{\det\bA} \ , \nonumber \\
\bA_{\alpha\beta}&=&\partial_\alpha
x^M\partial_\beta x^N
G_{MN}+2\pi\mF_{\alpha\beta}  \ , \nonumber \\
  \mF_{\alpha\beta}&=&
\partial_\alpha A_\beta-\partial_\beta A_\alpha-
(2\pi)^{-1}B_{MN}\partial_\alpha
x^M\partial_\beta x^N
 \ , \nonumber \\
S_{WZ}&=& \tau_p\int e^{2\pi i\mF}
\wedge C \ .  \nonumber \\
\end{eqnarray}
Our goal is to analyze description of the Wilson lines in ABJ
theory in terms of objects in supergravity dual of $AdS_4\times
CP_3$ with $B_{\rm{NS}}$ holonomy.

Let us start with the situation of Wilson loop description using
fundamental string. In fact, it turns out that fundamental string
description of Wilson loop is the same as in case of ABJM theory
and hence these strings describe symmetric combinations of Wilson
loops in fundamental representation. We will see below that the
effect of the non-zero parameter $b$ is sub-leading in
supergravity description. Equivalently, since the string with
world-volume topology AdS$_2$ is stretched in AdS$_4$ completely
it is clear that its dynamics is not affected by NS-NS two form
field.

Now we consider the Wilson loop in symmetric representation. In
case of $AdS_5/CFT_4$ correspondence these objects are described
by D3-branes that span $AdS_2$ subspace of $AdS_5$ together with
some $\Sigma_2$ subspace of $S^5$
\cite{Gomis:2006sb,Gomis:2006im}. In case of ABJM theory the
corresponding object is D2-brane that spans the subspace
$AdS_2\times S^1$ inside $AdS_4\times CP^3$. However since in this
case the D2-brane wraps $S^1$ in $CP^3$, it is clear that the
presence of two form $B_{NS}$ holonomy does not affect this
configuration at all.

Let us now consider the final example of D6-brane with topology
$AdS_2\times\Sigma_4$ where $\Sigma_4 \subset CP^3$. According to
standard arguments these D6-branes should correspond to Wilson
loops in anti-symmetric representation. More precisely we consider
D6-brane that spans $AdS_2$ subspace of $AdS_4$ and couples to
$C_{5}$. The similar calculation with zero $B_{NS}$ holonomy has
been performed in \cite{Drukker:2008zx} and we closely follow this
approach. Firstly, it is easy to see that the WZ term takes the
form
\begin{eqnarray}
S_{WZ}=\tau_6\int 2\pi F\wedge C^{(5)}
\end{eqnarray}
since the contribution $B\wedge C^{(5)}$ vanishes
due to the fact that D6-brane is extended in $AdS_4$ as well.
 On the other
hand the pullback of $B$ field will
appear in the DBI action. More
precisely, let us consider following
embedding ansatz
\begin{eqnarray}\label{ansD6}
\xi^1&=&\rho \ , \quad \xi^2=\psi \ ,
\quad  \
\xi^3=-\frac{1}{2}\chi \ ,
\quad \xi^4=\varphi_1 \ , \quad \xi^5=\theta_1 \ ,
\quad
\xi^6=\varphi_2 \ , \quad \xi^7=\theta_2 \ ,
 \nonumber \\
F_{\rho\psi}&=&-F_{\psi\rho}=f(\rho) \ , \quad u=0  \ ,
 \quad
\alpha=\mathrm{const} \ .
\nonumber \\
\end{eqnarray}
For this parameterization we obtain
following non-zero components of the
matrix $\bA\equiv \bA_0+B$
\begin{eqnarray}\label{bAex}
\bA_{\rho\rho}&=& \tR^2 \ , \quad
\bA_{\rho\psi}=-\bA_{\psi\rho}=2\pi
f \ , \quad  \bA_{\psi\psi}=\tR^2\sinh^2\rho \ , \nonumber \\
\bA_{33}&=&
4\tR^2\sin^2\frac{\alpha}{2}
\cos^2\frac{\alpha}{2}
\ , \nonumber
\\
\bA_{34}&=&\bA_{43}=
-2\tR^2\sin^2\frac{\alpha}{2}
\cos^2\frac{\alpha}{2}\cos\theta_1
 \ , \nonumber \\
\bA_{36}&=&\bA_{63}=
2\tR^2\sin^2\frac{\alpha}{2}\cos^2\frac{\alpha}{2}
\cos\theta_2 \ , \nonumber \\
\bA_{44}&=&
\tR^2
[\cos^2\frac{\alpha}{2}\sin^2\theta_1
+\sin^2\frac{\alpha}{2}\cos^2\frac{\alpha}{2}
\cos^2\theta_1] \ , \nonumber \\
\bA_{45}&=&-\bA_{54}=
-\frac{b}{4}\cos^2\frac{\alpha}{2}\sin\theta_1
\ ,
\nonumber \\
\bA_{46}&=&
-\tR^2\sin^2\frac{\alpha}{2}\cos^2\frac{\alpha}{2}
\cos\theta_1\cos\theta_2 \ , \nonumber
\\
\bA_{55}&=&
\tR^2\cos^2\frac{\alpha}{2} \ , \quad
 \bA_{77}=
\tR^2\sin^2\frac{\alpha}{2} \  ,
\nonumber \\
\bA_{66}&=&
\tR^2[\sin^2\frac{\alpha}{2}
\sin^2\theta_2+\sin^2\frac{\alpha}{2}
\cos^2\frac{\alpha}{2}\cos^2\theta_2] \
, \nonumber \\
\bA_{67}&=&
-\frac{b}{4}\sin^2\frac{\alpha}{2}\sin\theta_2=-\bA_{76}
\ ,  \nonumber \\
\end{eqnarray}
where $\bA_0$ are components of the matrix $\bA$ that do not
depend on $b$ and where $B$ are the matrix components proportional
to $b$.

In principle we can then evaluate
$\det\bA$. However the resulting
expression is very complicated and is
not very interesting. For that reason
we simplify the analysis using the
following observation. We know that  $b
\in [0,1)$. Then since $g$ is
proportional to $\tR^2$ we can presume
that $ (\bA_0)_{ij}B^{jk} <<
(\bA_0)^k_{i}$
 and
 hence we can write
\begin{eqnarray}\label{detbA}
\sqrt{\det \bA }
\approx\sqrt{\det
\bA_0}\left(1-\frac{1}{4}
\bA_0^{jk}B_{kl}\bA_0^{lm}B_{mj}\right)
\ .
\nonumber \\
\end{eqnarray}
Then the form of the ansatz (\ref{ansD6})
implies
\begin{eqnarray}
\det \bA_0=
4 \tR^{10} \sin^2\theta_1\sin^2\theta_2
\sin^6
\frac{\alpha}{2}\cos^6\frac{\alpha}{2}
(\sinh^2\rho+\beta^2 f^2) \
\nonumber \\
\end{eqnarray}
and
\begin{eqnarray}
B_{ij}\bA_0^{jk}B_{kl}\bA_0^{li}=
2B_{45}\bA_0^{55}B_{54}\bA_0^{44}
+2B_{67}\bA_0^{77}B_{76}\bA_0^{66} \ ,
\nonumber \\
\end{eqnarray}
where
\begin{eqnarray}
\bA_0^{55}&=&
\frac{1}{4\tR^2\cos^2\frac{\alpha}{2}}
 \ , \quad
\bA_0^{44}=
\frac{1}{\tR^2}\frac{1}{\sin^2\theta_1
\cos^2\frac{\alpha}{2}} \ ,
\nonumber \\
\bA_0^{66}&=&
 \frac{1}{\tR^2
 \sin^2\frac{\alpha}{2}\sin^2\theta_2}
\ , \quad
\bA_0^{77}=
\frac{1}{4\tR^2
\sin^2\frac{\alpha}{2}} \ , \nonumber \\
\end{eqnarray}
and where the explicit form of the of
$B_{ij}$ follow from (\ref{bAex}).
Using these results we finally find
that (\ref{detbA}) is equal to
\begin{eqnarray}
\sqrt{\det \bA }
\approx \tR^{5} \sin\theta_1\sin\theta_2
\sin^3
\frac{\alpha}{2}\cos^3\frac{\alpha}{2}
\sqrt{\sinh^2\rho+\beta^2 f^2}
\left(1+\frac{b^2}{64\tR^4}\right) \ .
\nonumber \\
\end{eqnarray}
Consequently the DBI part of D6-brane
action takes the form
\begin{eqnarray}
S_{DBI}&=&
\tau_6 \tR^7 \frac{k}{2\tR} \int
2\sin\theta_1\sin\theta_2
\sin^3\frac{\alpha}{2}\cos^3\frac{\alpha}{2}
\sqrt{\sinh^2\rho+\beta^2 f^2}
\left(1+\frac{b^2}{64\tR^4}\right)= \nonumber \\
&=&\tau_6 \pi^3 \frac{R^9}{2^3 k^2}
\int d\rho d\psi
\sin^3\frac{\alpha}{2}\cos^3\frac{\alpha}{2}
\sqrt{\sinh^2\rho+\beta^2f^2}
\left(1+\frac{b^2}{64\tR^4}\right)
 \ ,
\nonumber \\
\end{eqnarray}
where we have introduced the parameter $\beta$ defined as
\begin{equation}
\beta=\frac{8\pi k}{R^3}
\end{equation}
and where we have performed the following integration
\begin{equation}
\int_0^{2\pi}d\varphi_1
\int_0^{2\pi}d\varphi_2
\int_0^{2\pi}d\chi^3 \int_0^\pi
d\theta_1 \sin\theta_1 \int_0^\pi
d\theta_2\sin\theta_2= 2^5\pi^3 \ .
\end{equation}
Considering the WZ term we find
\begin{eqnarray}\label{SWZans}
S_{WZ}&=&i\tau_6\int 2\pi F
\wedge C^{(5)}=\nonumber \\
&=&\frac{i\tau_6 R^9}{2^{10}k^2} \beta
\int 2 f (\sin^2\alpha\cos\alpha+
2\cos\alpha-2)\sin\theta_1\sin\theta_2
\ .
\nonumber \\
\end{eqnarray}
Finally we perform the integration
 over
five coordinates of CP$^3$ in (\ref{SWZans}) that again
 gives  the factor $2^5\pi^3$.
Consequently the full D6-brane action takes the form
\begin{eqnarray}\label{S6final}
S_{DBI}+S_{WZ}&=&\frac{R^9\pi^3
\tau_6}{2^3  k^2} \int d\rho d\psi
[\sin^3\alpha \sqrt{\sinh^2\rho+\beta^2
f^2} \left(1+\frac{b^2k^2}{4R^6}\right)
+\nonumber \\
&+&i\beta
 f (\sin^2\alpha\cos\alpha+
2\cos\alpha-2)] \ .
\nonumber \\
\end{eqnarray}
Using this form of the action we easily
derive the equation  of motion for
$\alpha $
\begin{eqnarray}\label{eqalpha}
3\sin^2\alpha\cos\alpha
\sqrt{\sinh^2\rho+\beta^2 f^2}
(1+\frac{b^2k^2}{4R^6}) - 3i\beta
f(\sin^3\alpha)=0
\ .
\nonumber \\
\end{eqnarray}
Further the equations of motion for the
components of the gauge fields
$A_1,A_2$ imply an existence of
conserved electric flux $\Pi$
\begin{eqnarray}\label{Pi}
\sin^3\alpha\frac{\beta^2 f}{
\sqrt{\sinh^2\rho+\beta^2f^2}}(1+\frac{b^2k^2}{4R^6})
+i\beta
(\sin^2\alpha\cos\alpha+2\cos\alpha-2)=\Pi
\end{eqnarray}
that allows us to express $f$ as a
function of $\Pi$ so that
\begin{equation}
f=
\frac{(\Pi-i(\sin^2\alpha\cos\alpha+2\cos\alpha-2))\sinh\rho}
{\sqrt{\beta^4\sin^6\alpha(1+
+\frac{b^2k^2}{4R^6})^2
-\beta^2(\Pi-i(\sin^2\alpha\cos\alpha+2\cos\alpha-2))^2}}
\ .
\end{equation}
Finally, using (\ref{eqalpha}) we find
the dependence of $\Pi$ on $\alpha$
\begin{eqnarray}
\Pi=2i\beta(\cos\alpha-1)-i\beta
\sin^2\alpha\cos\alpha\frac{b^2 k^2}{2R^6} \ ,
\nonumber \\
\end{eqnarray}
where we neglected $\frac{b^4 k^4}{64
R^{12}}$. It is well known that the
electric flux is proportional to the
number of  dissolved fundamental
strings on the world-volume of
D6-brane. Since the electric flux is
aligned along $\rho,\psi$ directions we
find that the number of strings that
are stretched in these directions and
that are  localized at given $\alpha$
is equal to
\begin{eqnarray}\label{Q}
Q
= \tau_6 \frac{R^9}{8k^2}
\frac{i\Pi}{2}
=\frac{N}{2}\left[(1-\cos\alpha)+\frac{1}{2}
\sin^2\alpha\cos\alpha
\frac{b^2 k^2}{2 R^6}\right] \ , \nonumber \\
\end{eqnarray}
where the factor $i$ was included as a
consequence of the fact that we work in
space with Euclidean signature and
where the factor $2^5\pi^3$ follows
from the integration over $CP^3$.
Observe that now the number of
fundamental string is sensitive to the
presence of $B_{NS}$ holonomy.
%
%
Finally we determine the value of the
action (\ref{S6final}) for the ansatz (\ref{ansD6}) and
we find
\begin{eqnarray}\label{Sans}
S
=\frac{\pi^3 \tau_6 R^9}{8 k^2} \int
d\rho d\psi \sin \rho\left[
 \frac{\sin^2\alpha }
{\sqrt{1+\cos^2\alpha
\frac{b^2k^2}{2R^6}}}+
\frac{2\cos\alpha(\cos\alpha-1)}{
\sqrt{1+\cos^2\alpha
\frac{b^2k^2}{2R^6}}}\right]
\left(1+\frac{b^2k^2}{2R^6}\right)
\nonumber \\
\end{eqnarray}
However this is not full story since we
should perform a Legendre
transformation in this action in order
to write it as a functional of $\Pi$
instead of $f$ \cite{Drukker:2005kx}.
To do this we add to it an expression
\begin{eqnarray}\label{deltaS}
\delta S&=&
-\frac{\pi^3 \tau_6 R^9}{ 2^4k^2}
\int d\rho d\psi (\Pi f)=
\nonumber \\
&=&-
\frac{\pi^3 \tau_6 R^9}{ 2^4 k^2}
\int d\rho d\psi \sinh\rho
\left[\frac{2\cos\alpha(\cos\alpha-1)}
{\sqrt{1+\cos^2\alpha\frac{b^2k^2}{2R^6}}}
+\frac{\sin^2\alpha\cos^2\alpha\frac{b^2k^2}{2R^6}}
{\sqrt{1+\cos^2\alpha\frac{b^2k^2}{2R^6}}}\right]
\left(1+\frac{b^2 k^2}{2R^6}\right) \ .
\nonumber \\
\end{eqnarray}
Consequently the sum of  (\ref{Sans})
 with (\ref{deltaS})
gives
\begin{eqnarray}\label{Sfin}
S_{L.T.}=S+\delta S=
\frac{\pi^3 R^9\tau_6 }{ 2^4 k^2}
\int d\rho d\psi \sinh\rho
\sin^2\alpha
\left(1+\cos^2\alpha\frac{b^2k^2}{4R^6}\right)
\left(1+\frac{b^2 k^2}{2R^6}\right) \ .
\nonumber \\
\end{eqnarray}
To proceed further we use (\ref{Q}) to
express $\cos\alpha$ as a function of
$Q,N$ and $b$
\begin{eqnarray}
\cos\alpha=\frac{N-2Q}{N}
\left(1+\frac{b^2k^2}{4R^6}\right)
\nonumber \\
\end{eqnarray}
and following \cite{Drukker:2008zx} we
regularize the integration over $AdS_2$
in (\ref{Sfin}) with the result $[\int
d\rho d\psi \sinh\rho]_{reg}=-2\pi$.
Then using the relations
\begin{equation}
N=\lambda k \ ,
R^9=2^7\pi^3 k^3 \sqrt{2}\lambda^{3/2}
\end{equation}
we find that (\ref{Sfin})  is equal to
\begin{eqnarray}\label{Sfin1}
S_{L.T.}&=&-\pi
\sqrt{2\lambda}\frac{Q(N-Q)}{N}
\left(1+\frac{b^2
k^2}{4R^6}\left(1-\frac{N}{2Q}\right)\right)\times
\nonumber \\
&\times&  \left(1+\frac{N-2Q}{N-Q}
\frac{b^2k^2}{8R^6}\right)\left
(1+\frac{b^2 k^2}{4R^6}
\left(\frac{N-2Q}{N}\right)^2\right)
\left(1+\frac{b^2 k^2}{2R^6}\right) \ .
\nonumber \\
\end{eqnarray}
It is easy to see that (\ref{Sfin1}) is
invariant under exchange $Q\rightarrow
(N-Q)$.
The fact that the action is invariant under exchange $Q\rightarrow
N-Q$ implies that this D6-brane describes Wilson loop in
anti-symmetric representation.

Now we would like to interprete these
results from the point of view of Wilson loops in
ABJ theory.
 Luckily
we can use the results derived in case
of ABJM theory
\cite{Drukker:2008zx,Rey:2008bh}. Let
us again restrict to the case of
circular Wilson line that should be
non-zero. As was derived in
\cite{Drukker:2008zx,Rey:2008bh} the
expectation values of these Wilson
loops in weak coupling limit are equal
to
\begin{eqnarray}
W_M&=&1+\frac{\pi^2
M^2}{k^2}-\frac{\pi^2M^2}{6k^2} \ ,
\nonumber \\
\overline{W}_N&=&1+\frac{\pi^2
N^2}{k^2}+
\frac{\pi^2 N^2}{6k^2} \   \nonumber \\
\end{eqnarray}
so that their symmetric linear
combination is equal to
\begin{eqnarray}
W^+&=& \frac{1}{2}(W_M+\overline{W}_N)=
 1+\pi^2 \lambda^2(
1+\frac{5}{6}\frac{b}{\lambda}-\frac{5}{12}\frac{b^2}{\lambda^2})
\ .
\nonumber \\
\end{eqnarray}
If we extrapolate these results into
the strong coupling limit when $\lambda
>>1$ and use the fact that
$b\in[0,1)$ we find that all terms proportional to $b$ are
sub-leading. However this result suggests that the dual
description of the symmetric combinations of Wilson loops in
fundamental representations is given by fundamental string since
as we argued above  it is not sensitive to the presence of
$B_{NS}$ holonomy.

In the same way we can consider an
anti-symmetric combinations of Wilson
loops in fundamental representation and
we find
\begin{eqnarray}
W^-=\frac{1}{2}(W_M-\overline{W}_N)=
\pi^2\lambda^2
\left[-\frac{1}{6}+\frac{\pi^2}{2\lambda}
+\frac{5\pi^2}{12}\frac{b^2}{\lambda^2}\right]
\ .
\nonumber \\
\end{eqnarray}
It is interesting that the
anti-symmetric combination of the
Wilson loops contain the even powers of
$b$ only. Unfortunately it is not known
the dual description of these Wilson
loops.

Let us now consider Wilson loop in
anti-symmetric representation. It was
shown in \cite{Drukker:2008zx} that
in case of ABJM theory the D6-brane description
of this object gives the result
\begin{equation}\label{Ddruk}
W=-\frac{p(N-p)}{N}\pi\sqrt{2\lambda} \ ,
\end{equation}
where the factor $\pi\sqrt{2\lambda}$ corresponds to the Wilson
loop calculated in fundamental representation. Note that the value
of Wilson line does not change when we replace $p$ with $N-p$
which is characteristic property of Wilson line in antisymmetric
representation. Further, it was argued in \cite{Drukker:2008zx}
that the appropriate interpretation of (\ref{Ddruk}) in the dual
ABJM theory is given  a linear combination of the Wilson loops in
anti-symmetric representations where the first one corresponds the
trace over $U(N)$ and the second one the trace over
$\overline{U(N)}$.
Let us now return to the case of ABJ theory with the gauge group
$\overline{U(N)}\times U(M)$. We claim that the result derived in
previous section correctly describes the Wilson loop in
anti-symmetric representation of the rank $Q$ since as we have
shown it is symmetric under exchange $Q\rightarrow N-Q$. We also
mean that it is very interesting that this result is sensitive to
the presence of non-trivial $B_{NS}$ holonomy even if the
corrections are sub-leading in supergravity approximation. We can
compare this fact with the calculations of the Wilson loops in
symmetric representation when we argued that their supergravity
description does not see the presence of $B_{NS}$ at all.
\\
\vskip .2in \noindent {\bf
Acknowledgements:} We would like to thank the referee for various constructive 
suggestions regarding our paper. The work of JK  was supported by the Czech
Ministry of Education under Contract No. MSM 0021622409.

\end{document}